# Advanced Life Peaked Billions of Years Ago According to Black Holes

David Garofalo

Department of Physics, Kennesaw State University, Marietta, GA, 30060 USA; dgarofal@kennesaw.edu

**Abstract:** The link between black holes and star formation allows for us to draw a connection between black holes and the places and times when extraterrestrial intelligences (ETIs) had a greater chance of emerging. Within the context of the gap paradigm for black holes, we show that denser cluster environments that led to gas-rich mergers and copious star formation were places less compatible on average with the emergence of ETIs compared to isolated elliptical galaxies by almost two orders of magnitude. The probability for ETIs peaked in these isolated environments around 6 billion years ago and cosmic downsizing shifted the likelihood of ETIs emerging to galaxies with weak black hole feedback, such as in spiral galaxies, at late times.

## 1. Introduction

To determine the prevalence of intelligent life in the Universe, and to understand the potential evolutionary track of our own species ([22, 38]), we should understand the challenges that may prevent the evolution of life on planets. The length of time that a technologically advanced civilization can sustain itself (e.g., [37]) is of particular interest for both applications. The well-known Drake equation ([11]) estimates the number of extraterrestrial intelligences (ETIs) in our galaxy that can communicate their presence to us, as given by

$$N = R f_p n_e f_l f_i f_c L$$

R is the number of stars forming per year; $f_p$ is the fraction of stars that have planets; $n_e$ is the number of planets per star where life could emerge; and $f_l$, $f_i$ and $f_c$ are the fraction of those planets on which life emerges, the fraction of that life that evolves to intelligence and the fraction of ETIs that becomes communicative over interstellar distances via its technology, respectively. The final factor L is the length of time that the ETI remains communicative. While the equation was developed to explore the chance of detecting life in our galaxy, it is of interest to explore the question more broadly. From the perspective of the Universe as a whole, is it possible to uncover when ETIs had the greatest chance of emerging and where?

Our understanding of the processes that determine where and when star formation peaks in the Universe has matured significantly, to the point where we can begin to explore more broadly the question of intelligence across space and time. The formation of black holes and their feedback on the rate of star formation are, crucially, both galaxy- and redshift-dependent. We reframe the Drake equation and show that isolated field elliptical galaxies around redshift 2 constitute the hub where ETIs had the greatest chance of evolving. If we assume that it took about 5 billion years for technologically advanced civilization to emerge, the peak era for the emergence of cognition in the Universe, therefore, occurred 6 billion years ago.

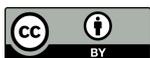

In Section 2, we describe the black hole–star formation connection. In Section 3, we reframe the Drake equation to the Universe at large. Additionally, in Section 4, we conclude the paper.

## 2. Black Hole Feedback on Stars across Space and Time

The connection between life and black holes is direct. Planets are associated with stars and stars are affected by black holes, as evidenced by the black hole scaling relations uncovered over two decades ago ([32], [19], [12]). Star and planet occurrence frequencies and formation in different galaxies must therefore relate to black holes and the nature of their feedback.

Since the 1970s, authors have quantitatively used accretion onto black holes to explain active galactic nuclei (AGN, e.g., [39]). After five decades, evidence suggests that different kinds of accreting black holes produce different kinds of feedback, with some black holes responsible for enhancing the star formation rate in AGN, while others are responsible for suppressing it ([26], [9]). Singh et al. [40] have combined the different types of black hole feedback into a unified picture, in which mergers that lead to counterrotating accretion disks around spinning black holes produce powerful, collimated jets of energy that enhance star formation rates. This is the so-called gap paradigm for black holes ([18]). Because counterrotation spins black holes down and then up again in the opposite direction, a prograde accreting phase is instantiated, which provides quantitatively different feedback.

In Figures 1 and 2, we describe the range of behavior for AGN with jets, so-called radio loud AGN, of which is related to star formation. Jetted AGN that have an effect on star formation emerge from mergers which trigger cold gas flows toward the center of a newly formed, rapidly rotating black hole, such that the disk of gas forms in counterrotation. In the left panel of both Figures 1 and 2, we have a drawing of clumps of gas coming in from the greater galaxy which eventually become reoriented such that they lie in a plane that is set by the direction of the angular momentum vector of the rotating black hole. This is the well-known general relativistic mechanism referred to as the Bardeen–Petterson effect ([4]). Counterrotation is associated with various general relativistic effects that maximize the power and collimation of the jet ([18]). This type of jet is funneled through the cold gas and pushes it into states of higher densities, thereby triggering star formation ([26]; [40]). However, there is a dependence on environment in the timescale for this enhancement. In isolated environments, the cold gas accretes onto the black hole at the maximum rate, the so-called Eddington accretion rate. This is given as

$$(dM/dt)_{Edd} = L_{Edd}/(\eta c^2)$$

where $L_{Edd}$ is the Eddington luminosity and $\eta$ is the efficiency of accretion. This allows for one to estimate a characteristic timescale for spinning the black hole down to a zero spin, a state in which the jets turn off. This is about 8 million years. In denser environments, instead, the black hole mass tends to be larger, the jet more powerful and its feedback effect greater. This has the effect of altering the state of accretion into an advection dominated accretion flow (ADAF) in a characteristic timescale of about 4 million years ([18]; [1]). ADAF accrete at least two-orders-of-magnitude lower than the Eddington rate, implying that the timescale to spin the black hole down to zero spin is slowed down by the same factor. As a result, on average, the richest environments produce powerful, collimated jets that enhance star formation for a timescale that is about two-orders-of-magnitude longer than in more isolated environments.

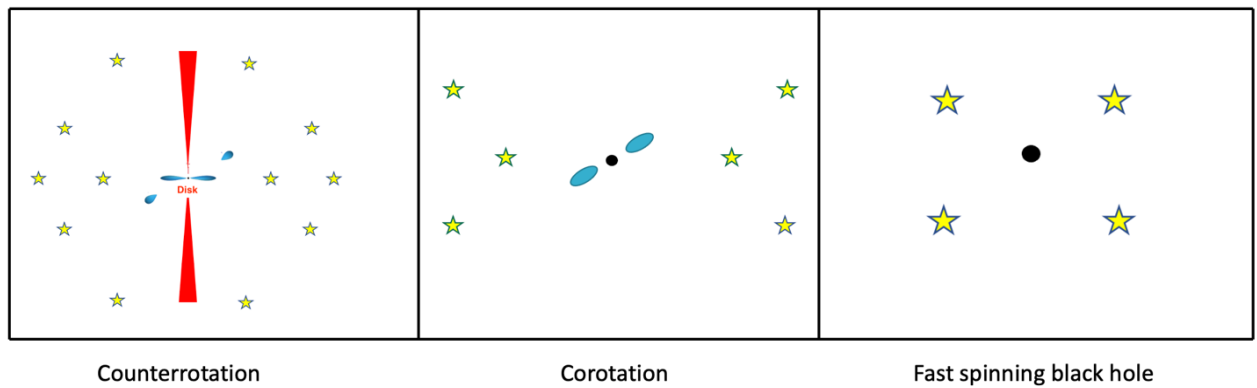

**Figure 1.** Isolated environments: Following a gas-rich merger, the accretion disk (in blue) settles in a counterrotating configuration around a rapidly spinning black hole. This configuration produces a narrow jet whose effect is to drive star formation rates to higher values (**left panel**). This enhancement lasts at most 8 million years after which the tilted disk fails to produce a jet, so star formation is unaffected (**middle panel**). The star formation rate dies down as the cold gas reservoir comes to an end and a dead quasar is formed after over a billion years (**right panel**).

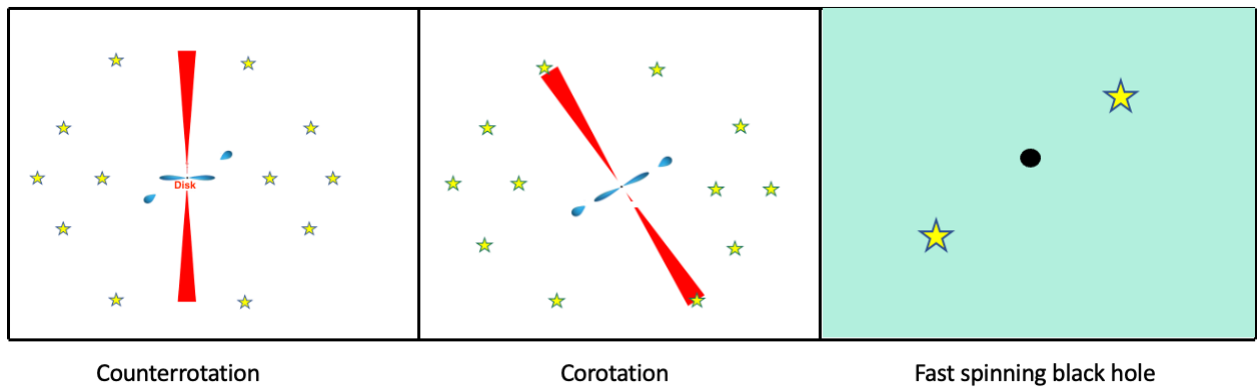

**Figure 2.** Richest environments: Following a gas-rich merger, the accretion disk (in blue) settles in a counterrotating configuration around a rapidly spinning black hole. This configuration produces a narrow jet whose effect is to drive star formation rates to higher values (**left panel**). This enhancement lasts for a few hundred million years after which the disk becomes tilted, a new jet direction is generated, and jet feedback suppresses star formation and heats the interstellar medium (ISM) with X-rays (**middle panel**). After billions of years, the gas supply dies down and the ISM of the dead quasar is suffused in a hot X-ray halo (**right panel**).

The crucial difference between the two types of environment is in the middle panel of both Figures 1 and 2, thus indicating the nature of the system as the black hole spins back up toward high values. This is simply the intrinsic evolution of a system that began in counterrotation and continues to accrete. Eventually, the black hole will be spun up to high spin values again by a disk that shares the same angular momentum direction as that of the black hole. The middle panel of both Figures 1 and 2 shows the accretion disk tilted with respect to the earlier counterrotating phase. This is due to the absence of the Bardeen–Pettersen effect at a zero black hole spin, such that the incoming gas forms a disk that maintains the angular momentum of the gas reservoir of the greater galaxy. The difference between the two middle panels, however, is that a new jet emerges only in the denser environment. This is due to the fact that ADAF accretion states do not suppress jet formation; the subtleties of this physical process can be found in the literature ([34], [18], [36]). The key difference is the presence of only positive AGN feedback in isolated environments, while both positive, then negative feedback, in richer ones. In the latter, in fact, as the black hole spins up to sufficiently large values compatible with a renewed powerful jet, the tilt makes the character of jet feedback different compared to the counterrotating phase. The tilt allows for the jet to directly impact the interstellar medium (ISM), heat it and shut down star formation (this is known as the Roy Conjecture, [14]). Even where stars do form planets, the negative feedback can limit habitability (e.g., [43]). The end state for such a system is a very low star formation and a hot X-ray halo permeating the galaxy. In more isolated environments, by contrast, stars evolve onto the main sequence undisturbed by AGN feedback.

### 3. The Drake Equation Viewed Broadly

We reframe the Drake equation so that it may apply to other galaxies and allow for us to treat the different kinds of black hole feedback. Two terms require attention. The first is the rate at which stars form and the second is the fraction of planets on which life can emerge. In order to determine these quantities, we refer to jet feedback in AGN explored in Singh et al. ([40]), and in particular, the evolution of radio loud AGN in the star formation–stellar mass plane. In the cases where counterrotation occurs, star formation is enhanced (captured in Figures 1 and 2) by the jet to a degree that, on average, is dependent on the environment. To understand the feedback on the rate of star formation, we pay attention to the green paths (as 'FRII phase') of Figure 5 in ([40]), showing the average behavior of AGN triggered in different environments by counterrotating black holes, which involves an enhancement in the star formation rate for the duration of the counterrotating phase. These counterrotating phases are followed, as previously described, by corotating accretion phases, which are in pink (labelled FRI phase). Of crucial importance is the fact that pink paths are smaller in isolated environments compared to rich environments. This is the key to understanding this work. The pink paths are associated with negative AGN feedback on the rate of star formation, so we focus on environments where the pink paths are the shortest. This leads us to consider isolated environments where star formation rates are, on average, enhanced but not suppressed. Such systems do not suppress star formation because they do not experience continued jet formation after the counterrotating phase (middle panel of Figure 1). Because counterrotation tends to be unstable, it is an unlikely configuration, which explains why jetted AGN are the minority (about 1 in 5 AGN has a jet—[8]). Isolated AGN constitute a fraction of all AGN, and isolated AGN that experience the green counterrotating FRII phase constitute a subset of those AGN. We are interested in a subset of those, i.e., in the extreme jetted AGN in isolated environments that do not experience pink paths because of a relatively weaker degree of jet feedback on their interstellar medium.

From [40], we obtain the peak in the rate of star formation by evaluating the highest point on the green paths for the different environments. As seen in their Figure 5, for isolated environments, we obtain a peak star formation rate of 15 solar masses per year and we compare this to the 3 solar masses per year for the Milky Way, which gives us an enhancement of 5. For denser group environments, we estimate a star formation peak rate of 40 solar masses per year, which divided by 3, gives us a 13.3 enhancement rate with respect to the Milky Way. For clusters, we have 200 solar masses and a relative enhancement factor of 66.7 with respect to the Milky Way. These values are shown in Table 1. The reason for these large star formation rates is that FRII jets enhance star formation rates; this kind of positive feedback is increasingly more dominant in denser environments.

**Table 1.** Relative star formation rates and dust content for galaxies that experience the most positive and/or negative black hole feedback. The last column is the overall contribution to the reframed Drake equation relative to the Milky Way.

| Type | SFR | R | Dust Mass | $\mu_d$ | $F_{Drake}$ |
|---|---|---|---|---|---|
| Clusters | 200 | 66.7 | $7 \times 10^4$ | $2.2 \times 10^{-3}$ | 0.15 |
| Groups | 40 | 13.3 | $2 \times 10^5$ | $6.2 \times 10^{-3}$ | 0.08 |
| Fields | 15 | 5 | $2.6 \times 10^8$ | 8.125 | 40.63 |
| Milky Way | 3 | 1 | $3.2 \times 10^7$ | 1 | 1 |

While star formation rates are enhanced in all these environments for this subclass of AGN with jets, the denser environments, again, tend eventually to be subject to star formation suppression due to tilted jets (Figure 2). The star formation rates drop to low values, as indicated by the extended pink, FRI Phases. Not only is much of the gas no longer forming stars, but the interstellar medium is rapidly becoming suffused by X-rays, which affects the chemistry on planets. In order to quantitatively determine the effect of this feedback on the ability of a planet to sustain life in this environment, we use dust content as a simple proxy. Specifically, we use estimates of the total dust content in representative samples of AGN in rich and isolated environments ([42]) and compare them to the dust content of the Milky Way. In other words, we divide the total dust content in a representative galaxy by the total dust content of the Milky Way and refer to the parameter as $\mu_d$. Overall, the reframed Drake equation receives a modification term referred to as $F_{Drake}$ in Table 1, which multiplies R by $\mu_d$. It tells us where in the Universe the chance of detecting advanced life is greatest. The answer is in isolated field environments.

Habitability has been shown to be a complex interplay of various processes. However, counterrotation between black holes and accretion disks tends to require massive black holes (i.e., [27], [17]). Although these isolated field environments are not the most massive, Figure 5 of [40] shows us that they inhabit regions of the SFR-SM plane with SM values above log SM = 10 and with SFR values above approximately log SFR = −1. As a result, they are potentially compatible with the habitability parameters determined in [10] (see also [41] for a more recent exploration of the parameter space).

We have honed in on the places *where* advanced life has a greater chance of emerging, but we have not addressed the issue of *when* this occurs. Counterrotating accreting black holes are the product of mergers and the merger function experiences its peak at a redshift of 2 (e.g., [5]). This, therefore, is the redshift corresponding to when the greatest number of isolated field galaxies experienced a merger that led to cold gas flowing into the nucleus of the newly formed galaxy and settling into counterrotation around the newly formed black hole. A redshift of 2 corresponds to 2.8 billion years after the Big Bang, which is equivalent to 11 billion years ago. However, the onset of the AGN was not the onset of life. Instead, the AGN triggered star formation, and with it, planet formation. On Earth, life emerged and evolved over billions of years, and it has taken ~4.5 billion years for human life to become capable of interstellar communication. Thus, we assume a fiducial value of 5 billion years, which brings us to 7.8 billion years after the Big Bang, or 6 billion years ago (see [6] for an estimate regarding the Milky Way). The difficulties associated with the detection of signals from such enormous distances are, of course, compounded compared to the Milky Way. However, we are able to detect the radio frequencies from synchrotron radiation produced by the jets in these active galaxies, so it is not unreasonable to search by way of SETI.

Not all triggering is associated with mergers. As the merger rate decreases, eventually secular processes begin to dominate. This is thought to occur around z = 1. While this does not lead to the large star formation rates and quasars triggered by mergers, it nonetheless is not associated with the strong negative feedback described above in the densest merger environments, so the rates of star formation should be relatively high. Therefore, the likelihood for the emergence of extraterrestrial life at these late times shifts to disk-like galaxies, such as the Milky Way, where the feedback from black holes may have a different impact (e.g., [29]). Minor mergers also begin to matter more at late times, reheating the halo gas and leading to an intermittent in-fall of gas and the fueling of star formation (e.g., [33]). However, secular processes and minor mergers are thought to be processes associated with the star formation main sequence. We are focused on a subgroup of isolated field galaxies that are the product of mergers and that also have the unlikely feature of counterrotation between the black hole and accretion disk at their heart. The SFR associated with these systems can reach values that exceed the star formation main sequence and have a large enough stellar mass that they can dilute the negative supernova feedback effect on habitability.

Finally, we note that these isolated elliptical galaxies are not expected to harbor low metallicities because they are AGN-triggered by mergers with abundant cold gas, possibly from a disk-like galaxy, so they are not the old, red-and-dead ellipticals, but objects with abundant cold gas. In fact, X-shaped radio galaxies, which are prescribed by the model ([16]) and observed ([25]) to be triggered in isolated environments, show strong emission line signatures, suggesting abundant cold gas accretion in the galactic nucleus. These objects are not like elliptical galaxies in general, which have older stellar populations and thus dominated by M stars whose planets' habitable zones are closer to the star and subject to stellar flares and tidally locked rotation, which work against the development of life. There are many other issues that we have not addressed (e.g., [29]). While we have some idea of where star formation is enhanced by the jets produced by black hole feedback, we have not addressed whether such regions have other features necessary for the production of planets where life could be sustained (e.g., [20], [28]). Recent work has shown that black hole feedback by accretion disk winds could hamper the habitability conditions by the flux of high energy radiation ([13]) as far removed from the black hole as 1 kpc ([3]) and especially near the galaxy core ([7], [44], [31], [35]), but see [23]). We do not explore this in detail but limit ourselves to highlighting that accretion disk winds and the locations where FRII jets push cold gas into higher densities are not directly connected since the FRII jet enhances the star formation rate in a funnel-like region centered on the black hole, while the disk winds occur further out from the disk. It is also important to notice from theory that the winds from jetted quasars are not as strong as the winds from non-jetted quasars, which may affect the galaxy as a whole ([2]).

## 4. Conclusions

Our understanding of black hole feedback on galaxies has matured recently to the point where it can inform an exploration of life in the Universe. In this work, we have anchored our investigation to the idea that counterrotation between an accretion disk around a spinning black hole leads to the most effective AGN feedback. While this feedback is initially positive, in that it enhances the star formation rate, it may become negative; this tends to be increasingly the case for richer environments. We end up with the counterintuitive idea that major, gas-rich mergers that lead to the greatest star formation rates in the Universe are not the places where stable, life harboring planets tend to be formed. Instead, we followed the prescribed feedback which led us to isolated field environments where the star formation rate did not reach the maximum possible value of all AGN. Crucially, however, the lack of negative feedback led to an overall term in the reframed Drake equation that is close to two-orders-of-magnitude greater than in the richer environments. Given that such feedback is mediated by the merger function, which experienced a peak at redshift 2, we also identify the times when stars were most affected by black holes. Since habitability studies have advanced significantly, the Drake equation should be seen as a rule of thumb for understanding life in the Universe. Given the times and places identified for ETIs in this work, we expect SETI searches to require that signals come from Kardashev Type III civilizations (although see [24]). To the extent that we may someday speak of a peak era for the emergence of technologically advanced life in the Universe, our simplified exploration of the emergence of life in the context of AGN feedback indicates that such a time is in the past. We on planet Earth are, therefore, latecomers.

**Funding:**

This research received no external funding

**Data Availability Statement:**

All data is produced within the document.

**Conflicts of Interest:**

The author declares no conflict of interest.